# Intensity product-based optical sensing to beat the diffraction limit in an interferometer

Byoung S. Ham[1,2,*]

[1]School of Electrical Engineering and Computer Science, Gwangju Institute of Science and Technology, Gwangju 61005, Republic of Korea
[2]Quantum Lidar, Gwangju 61005, Republic of Korea
(May 28, 2024; e-mail:[*]bham@gist.ac.kr)

**Abstract**
The classically defined minimum uncertainty of the optical phase is known as the standard quantum limit or shot-noise limit (SNL) originating in the uncertainty principle of quantum mechanics. Based on SNL, the phase sensitivity is inversely proportional to $\sqrt{K}$, where K is the number of interfering photons or statistically measured events. Thus, using a high-power laser is advantageous to enhance sensitivity due to the $\sqrt{K}$ gain in the signal-to-noise ratio. In a typical interferometer, however, the resolution remains in the diffraction limit of the K=1 case unless the interfering photons are resolved as in quantum sensing. Here, a projection-measurement method in quantum sensing is adapted for an interferometer to achieve an additional $\sqrt{K}$ gain in resolution. For the projection measurement, the interference fringe of an interferometer can be Kth-powered to replace the Kth-order intensity product. To understand many-wave interference-caused enhanced resolution, several types of interferometers are numerically compared to draw corresponding resolution parameters. As a result, the achieved resolution by the Kth power to an N-slit interferometer exceeds the diffraction limit and the Heisenberg limit in quantum sensing.

## 1. Introduction

Optical sensing and metrology have been one of the most important research topics in modern science and technologies for precision measurements [1-13]. In optical sensing, high-precision measurements have been pursued in physics [1-3], chemistry [4,5], biology [6,7], medicine [7], and even semiconductor industries [8,9] and military services [10,11]. The classical phase sensitivity or resolution limit is known as the shot-noise limit (SNL), which originates in the uncertainty relation between photon number and phase [12-15]. In SNL, the phase sensitivity is proportional to $1/\sqrt{K}$, where K is the intensity order of interfering photons in an interferometer or statistically provided measurement events. To increase the signal-to-noise ratio (SNR), thus, a higher-order K probe light must be used. However, the demonstration of SNL for K>1 has not been reported yet in interferometer-based optical sensing and metrologies. Although many-wave interference in a Fabry-Perot interferometer (FPI) or grating-based spectrometer is a well-known technique for high-resolution spectroscopy, it is still limited to the K=1 case of SNL with high-end optical and electronic systems [16].

To beat SNL, quantum sensing has been developed [17-27]. In quantum sensing, superresolution [20] and supersensitivity [21,22] have been studied using nonclassical lights such as maximally entangled photons of N00N states [12,14] and squeezed lights [15]. The photonic de Broglie wave (PBW) of the N00N state is a good example of the superresolution satisfying the Heisenberg limit (HL) [23-25] overcoming SNL, where the supersensitivity is an independent issue [20]. However, N00N state-based PBWs suffer from inherently inefficient generation processes limited by nonlinear optics [20-24] and nonperfect fringe visibility for N>2 in an interferometer [20,25]. Squeezed states cannot be used for superresolution at all, even though it has been well adopted for supersensitivity in gravitational wave detection below SNL [15]. Moreover, quantum sensing is not compatible with classical sensors or metrologies. Most of all, quantum sensing is far behind to beat classical sensors for the phase sensitivity in a real world due to the limited order N of N00N states and the working condition of extremely noisy environments [26,27].

Here, the projection measurement in quantum sensing [25] is adapted to an interferometer for the intensity-product-based optical sensing to show interferometric SNL beating the diffraction limit or Rayleigh criterion limited by K=1. Satisfying coherence optics, the proposed method is inherently compatible with all interferometer-based sensors. The original projection measurement scheme is to split an interferometer's output port into identical K ports using nonpolarizing 50/50 beam splitters (BSs) [28]. Here, projection measurement aims to distinguish interfering photons in a post-detection manner, as originally understood in quantum sensing [17,25]. Thus, the role of divided fields in Fig. 1(a) (see X, not shown) is to post-determine the number of interfering photons contributing to SNL on a single-shot measurement basis [28]. For the projection measurement, the maximum number N can be ideally equal to the photon number of the input light. Regarding the first-order (K=1) intensity correlation of the Mach-Zehnder interferometer (MZI) output fields, no difference exists between single photons and continuous-wave light [29], as demonstrated with a single photon for the same fringe [28,30]. This equality between quantum and classical approaches satisfies K=1 [29].

## 2. Models

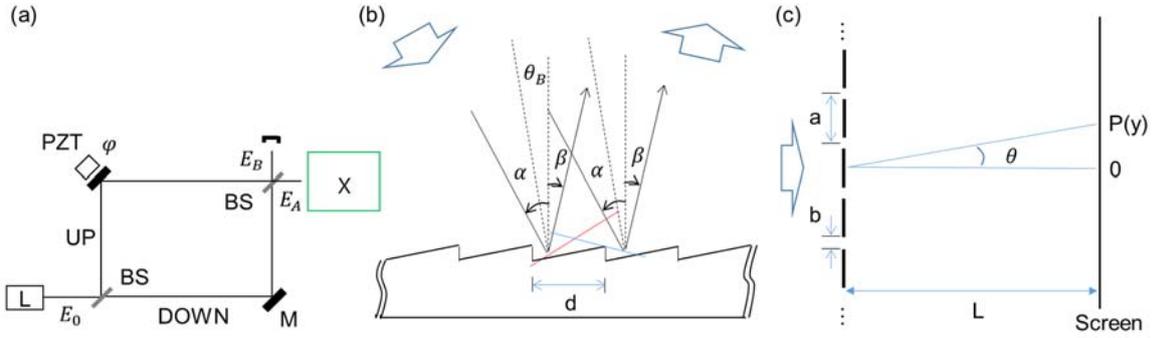

**Fig. 1.** Schematic of many-wave interference on a grating. Inset: N-slit interferometer. L: laser, BS: beam splitter, PZT: Piezoelectric transducer, X: projection measurements for the Kth power intensity product.

Figure 1 shows the schematic of the proposed projection measurement-based optical sensing using higher-order intensity correlations (products) for SNL. Figure 1(a) shows a typical MZI, and 'X' represents the projection measurement. For X, the MZI output is evenly divided into K ports by BSs for the Kth-order intensity correlation. Thus, 'X' is scalable by $2^K$, where K is the number of BSs. In the K-split MZI output ports, global phases generated by inserted BSs [32] and elongated optical paths to individual photodiodes do not affect their intensities due to Born's rule stating that measurement is the absolute square of its amplitude [31]. In other words, all K-divided output fields must be equal in intensity, satisfying the statistical ensemble of measurement events for SNL on a single-shot measurement basis.

For the coherence solution of the intensity-product-based optical sensing in Fig. 1(a), first, original MZI output intensities are derived as follows using the BS matrix [32]:

$$I_A = \frac{I_0}{2}(1 + cos\varphi), \tag{1}$$

$$I_B = \frac{I_0}{2}(1 - cos\varphi), \tag{2}$$

where $I_A = E_A E_A^*$, $I_B = E_B E_B^*$, and $E_j$ is the amplitude of the optical field. Thus, the MZI fringes show a deterministic coherence feature depending on the relative phase $\varphi$ caused by the MZI path-length difference. Due to the same coherence feature, MZI in Fig. 1 can be replaced by a Michelson interferometer, as usually adapted for remote sensors. Due to the global phase-independent intensities for all divided ports, the intensity of the jth divided output field in 'X' can be represented as:

$$I_j = \left(\frac{1}{2}\right)^K \frac{I_0}{2}(1+cos\varphi). \qquad (3)$$

Quantum mechanically, K implies interfering photons post-determined by the projection measurement: For a 1 W-power laser with 1 GHz bandwidth, the maximum photon number M is $10^9$. Using commercially available photodetectors whose response time is shorter than the inverse of the laser bandwidth, the Kth-order intensity correlation is as follows using Eq. (3):

$$C_N^{(K)} = \frac{I_0^K}{2^{(k+1)K}}(1+cos\varphi)^K, \qquad (4)$$

where $K \leq M$ and $K = \ln M/\ln 2$. Unlike quantum sensing using nonclassical light, the intensity product in Eq. (4) can be coherently amplified, compensating for the reduction factor of $2^{-(k+1)K}$. Unlike the enhanced coherence effect in many-wave interference (discussed below) [16], $I_0^K$ is the correlation effect by the intensity product. This correlation effect is powerful in reducing unwanted noise. Satisfying $K \ll M$, the intensity product in Eq. (4) gives a great benefit to the resolution of the proposed optical sensing with high SNR [28].

Figures 1(b) and (c) show schematics of many-wave interference on an N-groove grating and an N-slit interferometer, respectively. As introduced for MZI in Fig. 1(a), the interference fringe in Figs. 1(b) and (c) can also be used for the same projection measurement, satisfying SNL. In Figs. 1(b) and (c), the N-groove or N-slit resulting interference fringes show an enhanced resolution by $\Delta_N = \pi/N$ ($N \geq 2$), where Fig. 1(a) is only for N=2 [16]. For this, a discrete phase relation between N coherent waves is an essential requirement. Unlike the N-slit interferometer in Fig. 1(c), the N-groove grating in Fig. 1(b) allows only one interference fringe in each grating order due to the nearly equal ratio of 'a' to 'b.' Thus, the well-known grating equation is given by $2d\sin\theta_B = p\pi$ $(\frac{p\lambda}{2})$ for the grating order as well as for the ordered interference fringes. For the N-wave interference, the analytical solution can be derived from the N-slit interferometer in Fig. 1(c) [16]:

$$I_N(\alpha,\beta) = sinc^2\beta \left(\frac{sinN\alpha}{sin\alpha}\right)^2, \qquad (5)$$

where $\alpha = ka\sin\theta/2$, $\beta = kb\sin\theta/2$, and $k = 2\pi/\lambda$. The slit number N must be fully covered by the coherent input light. As discussed below, Eq. (5) results in N/2-enhanced resolution compared to the two-slit case of Eqs. (1) and (2) due to the N-wave superposition.

## 3. Results

Figure 2 shows numerical calculations of the Kth-order intensity correlations for Fig. 1(a) using Eq. (4). For this, the number of divided output ports is set at K=100, where K is far smaller than the actual photon number of $I_0$. All K-dependent intensity products are normalized for comparison purposes. As shown in Fig. 2, the ratio of full-width-at-half-maxima (FWHM) of the Kth-order to the first–order (K=1) intensity correlations is nearly $1/\sqrt{K}$ (see red circles in the right panel): The small discrepancy from the SNL theory (black curve) is due to the sine (monochromatic) function of the light rather than Gaussian distribution of the actual laser light (discussed in Fig. 3). Thus, the intensity product-resulting resolution enhancement in Fig. 2 demonstrates SNL for Eq. (4). In other words, Fig. 2 verifies that the proposed projection measurement-based optical sensing in MZI to satisfy SNL using a commercially available laser: For the experimental demonstrations up to N=4, see ref. 28. More importantly, the resulting $\sqrt{K}$-enhanced resolution in Fig. 2 is with no photon-number resolving single-photon detectors required in quantum sensing [12-14]. According to Born's rule, the K-divided output field's intensities in 'X' of Fig. 1 are independent and individual for measurements, satisfying the statistical ensemble of SNL. Due to the limited scalability by $2^K$ far less than the actual photon number of the laser, the phase-resolution enhancement is even practical due to high SNR as in conventional sensors.

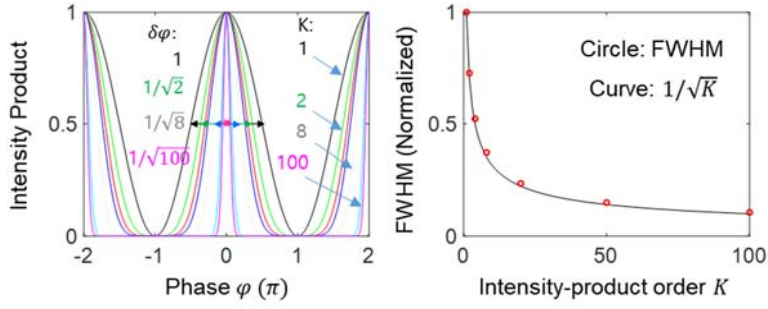

**Figure 2.** Numerical calculations of Eq. (3) for m=1(black), 2(green), 3(red), 4(blue), 8(dotted), 50(cyan), 100(magenta). The ordered intensity products are normalized. (right panel) curve: $1/\sqrt{K}$. Circles: data from the left panel.

Figure 3(a) shows numerical calculations of FWHMs of the autocorrelation (self-intensity product) for a Gaussian function. Here, a laser has the same feature as the Gaussian distribution if the mean photon number is $\langle n \rangle \gg 1$. Figure 3(b) is for linearly distributed fields for a comparison purpose. The horizontal axis in Figs. 3(a) and (b) is for the phase variation (noise) in the unit of standard deviation $\sigma$ of the Gaussian function used for Fig. 3(a). Figure 3(c) shows the ratio of the K-ordered FWHMs to the first order (K=1) for Figs. 3(a) and (b), where 'HL' represents the Heisenberg limit in quantum sensing as a reference. Thus, SNL is analytically confirmed for the Gaussian distribution of Fig. 3(a), where the FWHMs are inversely proportional to $\sqrt{N}$. This $\sqrt{N}$-enhanced phase sensitivity is due to the normal probability distribution of statistical events (see the blue dots in Fig. 3(c)). Thus, the origin of SNL in Fig. 2 is the Gaussian distribution of a laser. If the probability distribution is linear as shown in Fig. 3(b), the resolution enhancement is much better than the Gaussian (see the red dots in Fig. 3(c)). Thus, the enhancement factor in resolution can be higher if the photon distribution is non-Gaussian. Such an enhancement can be accomplished by frequency modulation as in a typical radar system [10,11]. Even in this case, the maximum sensing gain is still below the Heisenberg limit of quantum sensing, as shown by the gray curve in Fig. 3(c).

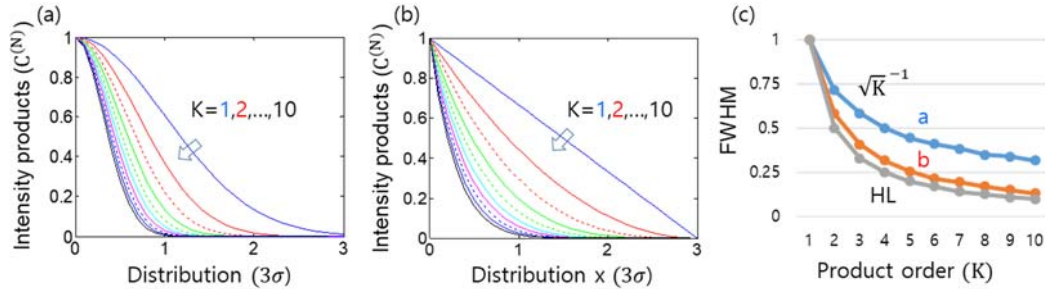

**Fig. 3.** Numerical calculations of intensity-order dependent FWHMs. (a) Gaussian. (b) Linear. (c) FWHM ratio vs. intensity-order. a: for (a); b: for (b). HL: Heisenberg limit (1/N). $C^{(K)} = (I_1)^K$.

Figure 4 shows numerical calculations of Eq. (5) for the many-wave interference in an N-slit system of Fig. 1(c). For the analysis, the slit number is set at $2 \leq N \leq 20$. As shown in Fig. 4(a), the fringe condition is satisfied by $\alpha = p\pi$, where $p = 0, \pm 1, \pm 2, ...$ As N increases, the fringe resolution gets better. To understand the N effect, N-dependent interference fringes are shown in Figs. 4(b) and (c), where N=2, 10, 40 are set for comparison purposes. For this, FWHM from Fig. 4(a) is calculated and plotted in Fig. 4(d). The red curve is the theoretical reference of $\pi/N$ [16]. At a glance, both numerical data from Fig. 4(a) and the reference seems to match well. The small discrepancy, however, is due to the non-Gaussian function of $I_N(\alpha, \beta)$ based on monochromatic waves as discussed in Fig. 3. Interestingly, this N-wave-caused resolution enhancement exactly the same as that in quantum sensing defined by the Heisenberg limit [12-14,20-25].

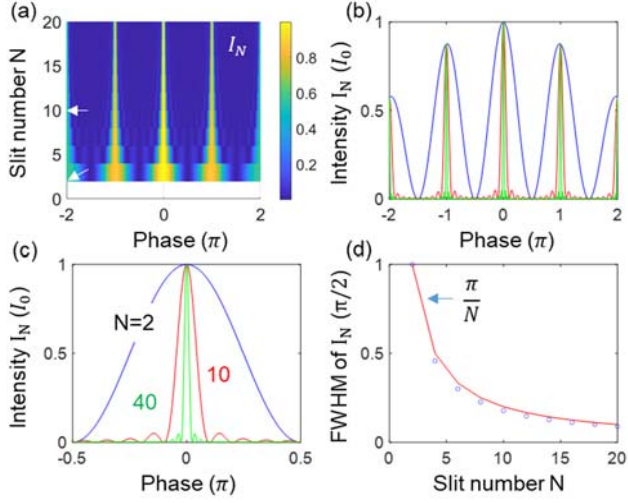

**Fig. 4.** Numerical simulations of N-slit interference. (a) First-order intensity correlation $I_N(\varphi)$. (b) and (c) blue (N=2), red(N=10), green (N=40). (d) Resolution (full-width at half maxima). Red curve: $\pi/N$. a=3b (see Fig. 1(c)).

Figure 5 shows numerical calculations of the Kth-order intensity products for Fig. 1(c). For this, two variables of N and K are set for $2 \leq N \leq 200$ and $1 \leq K \leq 40$, where N is far smaller than the actual photon number of $I_0$. All K-dependent intensity products $I_N^K$ are normalized for comparison purposes. As shown in Figs. 5(a)~(c), the ratio of FWHM of the Kth-order to the first–order intensity product is satisfied by $1/\sqrt{K}$ (see blue diamonds in the right panel), where the red curve is for the reference of SNL. The small discrepancy between them is due to the sine (monochromatic) function of the light rather than the Gaussian distribution of the actual laser light as discussed in Figs. 2 and 3. Thus, the intensity product-resulting resolution enhancement in Fig. 5(c) also demonstrates the same SNL in MZI in Fig. 2. In other words, Fig. 5 verifies that the proposed projection measurement-based optical sensing for the Kth power is effective to the measured intensity of an interferometer. Due to the practically unlimited Kth order, the phase-resolution enhancement can beat the Heisenberg limit given by $\pi/N$ (discussed in Fig. 6) [12-14].

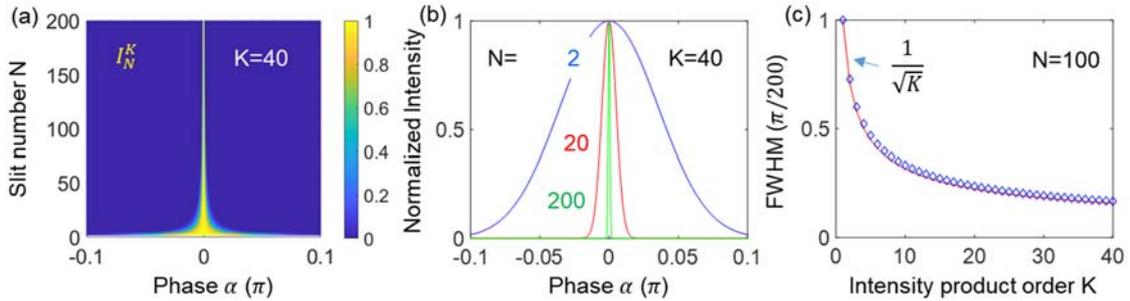

**Fig. 5.** Numerical simulations of Kth intensity correlation in an N-slit interferometer. (a) N=2~200, K=40. (b) for (a), where N=2,20,200. K=40. (c) diamonds: FWHM for N=100 in (a). Red curve: SNL.

Figure 6 shows a practical example of the grating-based spectrometer with N=1000. As discussed in Figs. 4 and 5, the resolution is enhanced by 500 times compared with the two-slit case. Figure 6(a) shows the interference fringe for Figs. 1(b) and (c) as a function of frequency of the probe light and phase difference α (or position P on a screen focused by a lens). The frequency $f_0$ is a reference, where an unknown frequency detected by an arrayed photodiode is calculated with respect to the position, i.e., the phase difference α. As demonstrated in Figs. 4 and 5, the phase resolution by 1000 slits or grooves in Fig. 6(a) is enhanced by π/1000.

Figure 6(b) shows the resolving power to separate an unknown frequency $f'$ from the reference $f_0$. The frequency difference of $f'$ from $f_0$ is easily calculated by measuring the detuned phase Δ from the principle maxima at α = −π. Here, the frequency $f'$ is chosen for $0.999f_0$ in Fig. 6(a). As shown in Fig. 6(b), $f'$ is resolvable by the Rayleigh criterion [16], where the N-enhanced resolution results in $\alpha = -\pi\left(1 + \frac{1}{1000}\right) = -3.1447$ and Δ= −0.001π.

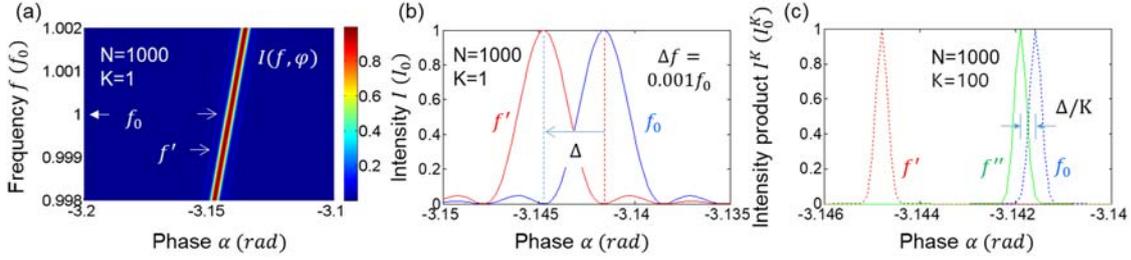

**Fig. 6.** Numerical calculations of N-slit interference. (a) Intensity $I(f, \alpha)$ for N=1000 and K=1. (b) Details of (a) for frequency resolution. (c) Enhanced frequency resolution with intensity-product order K=100.

If the proposed Kth power is applied to Fig. 6(a), then more resolvable frequencies are allowed as discussed in Fig. 5. For K=100, the resolution of $I_{1000}$ for N=1000 is 10 times enhanced to π/10,000, as shown by the blue and red dotted curves in Fig. 5(c). Thus, the unresolvable $f''$ in Fig. 5(b) by locating between $f_0$ and $f'$ is now resolvable, as shown by the green curve. This may sound awkward because we believe the measurement cannot retrospectively affect the optical system (interference). As the quantum eraser has been intensively studied over the last several decades for the mysterious phenomenon of the cause-effect relation [33-35], the enhanced resolution by the proposed intensity product method looks mysterious, too.

Regarding applications of the proposed intensity-product method to an interferometer such as an optical spectrometer, the measured interference signal can be Kth powered electronically to get an enhanced resolution as shown in Figs. 5 and 6. Achieving higher resolving power of an unknown optical signal has been the ultimate goal in optical spectroscopy over the last century. In both classical and quantum physics, the resolution of π/N is the ultimate limit to be achieved e by N amplitude superposition and intensity correlation, respectively. Thus, Fig. 6(c) is unprecedented and opens the door to a new field of ultra-resolution spectroscopy.

In Fig. 7, FPI, N-slit interferometer, and superresolution [12-14,36,37] are numerically investigated for the corresponding parameters of the resolution limit. The top (bottom) row is for less (more) dense cases with N. The left-end and middle-left columns are for FPI and N-slit cases [16]. The middle-right column is for the superresolution of quantum sensing [37]. The right-end column is for comparison between them. For FPI, the transmitted intensity is $I_T(r) = 1/[1 + (2r/(1-r^2))^2 sin^2\delta]$, where $\delta = 2kd$ is the phase gain between cavity mirrors at distance d, and r is the reflection coefficient of the cavity mirror. For superresolution [14], a typical MZI is reconfigured for the quantum eraser with orthogonal polarization bases [34,35], where the MZI output is divided into K folds for polarization-basis projection measurement through a polarizer [34-37], otherwise a single photon resolving detector must be used [14,38].

In the right-end column of Fig. 7, the resolution of FPI (N slit) is represented by the red (blue dotted) curve, representing the same FWHM. Thus, N=1000 (10,000) corresponds to r=0.999 (0.9999) in FPI, satisfying the same relation between N and r. Practically, however, FPI has a higher-order benefit with longer d, surpassing the grating-based spectrometer in achievable resolution. The N number in the middle-right column for superresolution represents the divided fields for intensity products. This intensity product is the same as SNL in Figs. 5 and 6, but the phase control of each divided field is required [37], resulting in N00N-based photonic de Broglie waves [20-24]. In the right-end column, superresolution is represented by the green curve, demonstrating the same resolution as the N-slit system and FPI. Unlike the projection measurement in superresolution and SNL, the FPI and N-slit interferometers are for amplitude superposition-based first-order intensity correlation, as shown in Eq. (5) for Fig. 4 [16]. As a result, either N wave superposition for the first-order intensity correlation or N projection measurement-based superresolution shows the same Heisenberg limit in resolution. This fact has never been discussed yet.

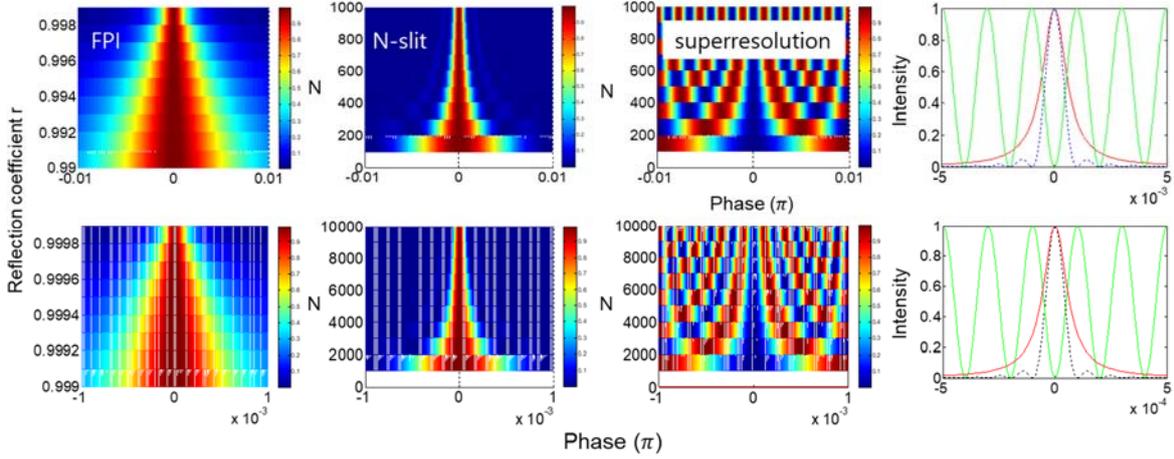

**Fig. 7.** Comparison between FPI and N-slit interferometer. (left-end) FPI. (middle-left) N-slit interferometer. (middle-right) Superresolution. (right-end) Fringes. red: FPI; dotted: N-slit; green: superresolution.

Unlike superresolution [37], the proposed intensity product-based SNL completely excludes phase relations between fields and thus satisfies the statistical ensemble of measurement events. As demonstrated in Hanbury-Brown and Twiss experiments [39], the advantage of the proposed intensity-product method over the amplitude interference in FPI is the phase-variation independence among fields due to the phase-independent identical intensities by the Born's rule [31]. This benefit has already been applied to optical spectrometers as well as quantum technologies using entangled photon pairs, even though the phase relation between entangled photons has still been veiled. Thus, SNL applied to the N-slit interference fringes in Fig. 6(c) surpasses the maximum resolution achievable by quantum sensors confined by the Heisenberg limit. This unprecedented resolution can be of course applied to conventional grating-based spectrometers or FPI-based wavelength meters.

The technical advantage of the proposed intensity product-based sensing method can also be found in a Si-photonics-integrated optical chip [40]. In general, the resolution of FPI strongly depends on the reflection coefficient of the cavity mirror, as shown in Fig. 7. The discrete N fields from an N-slit (grating) system are an extreme case of FPI due to the same amplitudes, resulting in the $\pi/N$ resolution (see the last column in Fig. 7), which is equal to the Heisenberg limit in quantum sensing [14]. However, FPI and the N-slit system are optically bulky and extremely sensitive to the phase variation caused by environments, i.e., temperatures, mechanical vibrations, and air turbulences. Thus Si-photonics can replace the bulky spectrometer for a robust micro-sensor beyond the diffraction limit. Moreover, the proposed sensing technique can also be applied for remote sensors such as a Doppler Lidar and hazardous gas detector to extend operational distance.

## 4. Conclusion

An intensity product-based optical sensing method was proposed to surpass the limited resolution in conventional spectrometers. For the projection measurements of interference fringes, SNL was satisfied not only for MZI but also N-slit interferometer. The N-slit interference was analyzed for the same resolution as superresolution in quantum sensing satisfying the Heisenberg limit. Due to the same physics of discrete phase relation between N waves, FPI was also compared with the N-slit interferometer, resulting in an equivalent parameter relation between reflection coefficient 'r' and slit number 'N.' Finally, the intensity product applied to the conventional spectrometer was numerically demonstrated for beating quantum sensors in resolution. Due to the satisfaction of the statistical ensemble by the proposed intensity-product sensing method, a simple electronic circuit for the Kth power of the interference fringes might be applied to the conventional spectrometer for the $\sqrt{K}$-enhanced resolution. Due to the coherence feature, a cw frequency modulation in the Radar technology might be useful for the proposed intensity products to extend operational distance.

**Funding:** This research was supported by the MSIT (Ministry of Science and ICT), Korea, under the ITRC (Information Technology Research Center) support program (IITP 2024-2021-0-01810) supervised by the IITP (Institute for Information & Communications Technology Planning & Evaluation). This work was also supported by a GIST research project granted by GIST in 2024.
**Data Availability:** All data generated or analyzed during this study are included in this published article.
**Conflict of Interest:** The author declares no competing interest.